\documentclass[aps,prb,reprint,superscriptaddress,floatfix,nofootinbib]{revtex4-2}

\usepackage[utf8]{inputenc}
\usepackage[T1]{fontenc}
\usepackage{lmodern}
\usepackage{graphicx}
\usepackage{amsmath,amssymb,amsfonts,bm}
\usepackage{braket}
\usepackage{xcolor}
\usepackage{hyperref}
\usepackage{booktabs}
\usepackage{xspace}

\newif\ifshowtodos
\showtodosfalse  

\ifshowtodos
  \newcommand{\TODO}[1]{{\color{red}\textbf{[TODO: #1]}}}
  \newcommand{\NOTE}[1]{{\color{blue}\textit{[note: #1]}}}
\else
  \newcommand{\TODO}[1]{}
  \newcommand{\NOTE}[1]{}
\fi

\newcommand{\hk}{\ensuremath{H(\mathbf{k})}\xspace}
\newcommand{\sk}{\ensuremath{S(\mathbf{k})}\xspace}
\newcommand{\Ef}{\ensuremath{E_{\mathrm{F}}}\xspace}
\newcommand{\slakonet}{\textsc{SlaKoNet}\xspace}
\newcommand{\jarvis}{\textsc{Jarvis}\xspace}

\hypersetup{
  colorlinks=true,
  linkcolor=blue!50!black,
  citecolor=blue!50!black,
  urlcolor=blue!50!black
}

\begin{document}

\title{SlaKoNet-VQD: A universal Slater-Koster tight-binding Hamiltonian for variational
       quantum band-structure calculations on near-term hardware}

\author{Akshaya Ajith}
\author{Jaehyung Lee}
\author{Charles Rhys Campbell}
\author{Kamal Choudhary}
\email{drkamal@jhu.edu}
\affiliation{Department of Materials Science and Engineering,
             Johns Hopkins University, Baltimore, MD 21218, USA}
\affiliation{Department of Electrical and Computer Engineering,
             Johns Hopkins University, Baltimore, MD 21218, USA}


\date{\today}

\begin{abstract}
Variational quantum algorithms such as the variational quantum eigensolver (VQE)
and variational quantum deflation (VQD) have emerged as leading candidates for
computing electronic structure on near-term quantum hardware. For periodic
solids, however, their reach has been limited not by the quantum algorithm
itself but by the cost of constructing a faithful second-quantized Hamiltonian
for each material, typically via density-functional theory (DFT) followed by
Wannierization or by hand-fitted empirical tight-binding parameters.
\slakonet, a recently developed method, significantly reduces the computational cost of electronic Hamiltonian construction with a small accuracy tradeoff. It integrates modern deep-learning methods and the Slater-Koster tight-binding formalism to fit thousands of onsite and 
distance-dependent hopping and overlap tight-binding parameters across $65$ elements of
the periodic table. After fitting, these parameters can be used to deterministically construct an electronic Hamiltonian for any crystal made up of any of the 65 elements in the training set, and the crystal's bandstructure can subsequently be computed. In this work, we combine a \slakonet model trained on the JARVIS-TBmBJ dataset with a Qiskit-based variational quantum deflation algorithm, replacing the traditional and costly Hamiltonian construction step with a universal neural Hamiltonian generator. Within the 65-element set, the resulting \slakonet-VQD workflow is structure-agnostic, fully differentiable, fast, and suited to high-throughput screening of bandstructures.
We benchmark the workflow on silicon, recovering the full eight-band band
structure along the standard high-symmetry $k$-path with a mean absolute
deviation of $1.78$~meV from exact diagonalization on a $3$-qubit statevector
simulator.
We further compute the full band structure for five additional conventional
superconductors, Al, Ta, Nb, V, and ZrN, achieving a mean absolute deviation
within approximately $1.8$~meV, and demonstrate full execution of \slakonet-VQD on IBM Quantum
hardware for a representative $k$-point ground-state calculation on aluminum
with a mean absolute error of roughly $0.37$ eV.
Beyond the single-particle level, we promote the same \slakonet Hamiltonian to
a correlated Hubbard model and solve it with dynamical mean-field theory,
recovering the weakening of correlations across the group-5 metals and
strong quasiparticle renormalization in the cuprate La$_2$CuO$_4$, and we
identify the resulting impurity problem as the natural target for a quantum
solver.
The pipeline opens a path to high-throughput benchmarking of variational
quantum algorithms across the periodic table and to gradient-based
co-optimization of ansatz and Hamiltonian for materials discovery. An interactive web app is available at website: \url{https://atomgpt.org/quantum}.
\end{abstract}

\maketitle

\section{Introduction}
\label{sec:intro}

Quantum computing has been heralded as a natural platform for electronic-structure
problems, with the variational quantum eigensolver
(VQE)~\cite{peruzzo2014, kandala2017} and its excited-state extension,
variational quantum deflation (VQD)~\cite{higgott2019}, emerging as the
leading near-term algorithms. While the molecular case has matured into a
well-mapped landscape~\cite{cao2019, mcardle2020}, periodic solids remain
substantially less developed. Recent work has begun to bridge this
gap~\cite{choudhary2021, sherbert2021, vqesolids2025, iopvqd2025, gaaspmc2025},
but the input to the quantum solver, a faithful second-quantized Hamiltonian,
has invariably been built by hand for each material. Two routes dominate:
density-functional theory (DFT) followed by Wannierization to obtain a localized
tight-binding (TB) Hamiltonian~\cite{marzari2012, pizzi2020}, or hand-fitted
empirical Slater-Koster parameters~\cite{slater1954}. Both are accurate when
properly executed, but both demand human expertise per material and are
difficult to compose into a high-throughput or differentiable pipeline.

This per-material setup cost has become the dominant bottleneck in scaling
quantum band-structure studies. Constructing a maximally localized Wannier
Hamiltonian requires careful selection of disentanglement
windows~\cite{souza2001}, initial projections, and convergence checks;
empirical TB models require fitting against a target band structure or
experimental data. Neither workflow is differentiable, neither transfers
across chemistries without re-fitting, and neither composes naturally with
the gradient-based optimization that underpins modern variational quantum
algorithms.

In parallel, machine-learning approaches to electronic-structure
Hamiltonians have matured to the point of replacing these hand-built
inputs. End-to-end graph neural networks such as
DeepH-E3~\cite{li2022deephe3}, HamGNN~\cite{gu2023hamgnn}, and equivariant
wavefunction predictors~\cite{unke2021se3} learn the Hamiltonian matrix
elements directly from the local environment;
\slakonet~\cite{choudhary2025slakonet} pursues a complementary route that
keeps the physically interpretable Slater-Koster (SK) tight-binding
parametrization and instead uses automatic differentiation to learn the
distance-dependent SK hopping and overlap integrals themselves, across
$65$ elements of the periodic table, against TBmBJ-level JARVIS-DFT
data~\cite{choudhary2018optb88vdwTBmBJ, choudhary2020}. Once trained,
\slakonet constructs the tight-binding Hamiltonian \hk and overlap matrix \sk for
any composition within its training domain by polynomial interpolation of
the learned parameters at runtime. Critically for our purposes, the entire
mapping from atomic coordinates to \hk is differentiable, runs on the GPU,
and requires no system-specific manual parametrization.

In this work we close the loop by feeding \slakonet's tight-binding Hamiltonian
directly into a Qiskit~\cite{qiskit2024} VQD pipeline. The resulting
\slakonet-VQD workflow is, to our knowledge, the first quantum
band-structure calculation in which the Hamiltonian is produced by a
pretrained universal TB model rather than by per-material DFT,
Wannierization, or hand-fitting. We make five specific contributions. First, we
build a clean pipeline: an interface between the \slakonet Hamiltonian generator and the Qiskit \texttt{Estimator} primitive, including a Hermitization step
that handles finite-precision numerical noise from the Hamiltonian generation step (Sec.~\ref{sec:methods}). Second, we provide a validation: a
full-Brillouin-zone benchmark on silicon along the standard high-symmetry path,
comparing 3-qubit VQD eigenvalues to exact diagonalization at every $k$-point
(Sec.~\ref{sec:si-bandstructure}). Third, we report hardware execution:
running on IBM Quantum hardware with error mitigation and quantifying the descrepancy between simulated hardware and real hardware in this context (Sec.~\ref{sec:hardware}). Fourth, we demonstrate
generalization: with no retraining, the same workflow produces band
structures for a panel of conventional superconductors spanning elemental
free-electron and transition metals and a transition-metal nitride
(Sec.~\ref{sec:multimaterial}). Fifth, we sketch a correlated extension,
promoting the same neural Hamiltonian to a Hubbard model solved by dynamical
mean-field theory, recovering known correlation trends across the group-5
metals and strong quasiparticle renormalization in the cuprate La$_2$CuO$_4$,
and identifying the resulting impurity problem as the natural target for a
quantum solver (Sec.~\ref{sec:dmft}).

The remainder of the paper is organized as follows. Section~\ref{sec:methods}
describes the \slakonet model, the Hermitian solver, the VQD ansatz, and the
hardware-execution pipeline. Section~\ref{sec:results} presents single-$k$-point
benchmarks, full band structures, hardware results, multi-material
generalization, and the correlated DMFT extension. Section~\ref{sec:discussion}
discusses what this enables for high-throughput quantum materials screening and
the limitations of the current implementation. Section~\ref{sec:conclusion}
closes with an outlook on phonons, the path to an ML-accelerated quantum
dynamical mean-field theory, and fault-tolerant quantum band-structure
calculations.

\section{Methods}
\label{sec:methods}

\begin{figure*}[t]
  \centering
  \includegraphics[width=\textwidth]{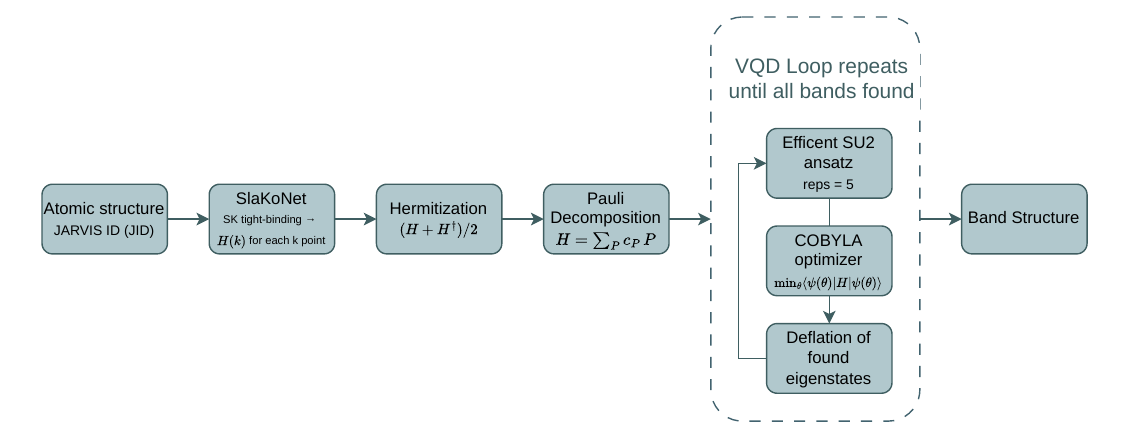}
  \caption{End-to-end \slakonet-VQD pipeline to determine full band structure. For a given material, \slakonet takes the atomic structure from JARVIS and produces the $k$-resolved tight-binding Hamiltonian $H(k)$ for each $k$-point, each of which is Hermitized via $(H + H^\dagger)/2$ then decomposed into the Pauli basis. The VQD loop (dashed box) is applied iteratively at each $k$-point. After each eigenstate is found, it is deflated from the spectrum so that the next application returns the following eigenvalue, repeating until all bands are recovered. The resulting eigenvalues are validated against exact diagonalization of the same Hamiltonian. }
  \label{fig:pipeline}
\end{figure*}

The complete workflow, from a crystal structure to validated quantum
eigenvalues, is summarized in Fig.~\ref{fig:pipeline} and detailed in the
subsections below.

\subsection{The \slakonet universal Slater-Koster framework}
\label{sec:methods-slakonet}

\slakonet~\cite{choudhary2025slakonet} is a parameter optimization framework
that produces a non-orthogonal tight-binding (TB) Hamiltonian and overlap in
the Slater-Koster (SK) two-center form~\cite{slater1954, porezag1995} for an
arbitrary periodic crystal containing elements with atomic number $Z\le 65$.
In contrast to graph-neural-network approaches that predict Hamiltonian
matrix elements as a function of the local environment, \slakonet directly
parametrizes the onsite energies and hopping and overlap integrals themselves. The model learns them
end-to-end via automatic differentiation against DFT-generated tuples of crystals and their calculated bandgaps and densities of states.

For a minimal $sp^3$ basis, the on-site energies $\varepsilon_s,\varepsilon_p$
and the SK hopping integrals $V_{ss\sigma},V_{sp\sigma},V_{pp\sigma},V_{pp\pi}$
(with corresponding overlap integrals) are stored as one-dimensional vectors
on a uniform interatomic-distance grid extending to $7~\text{\AA}$ about each atom. For
elements requiring $d$ orbitals, the basis is extended by
$\varepsilon_d, V_{sd\sigma}, V_{pd\sigma}, V_{pd\pi}, V_{dd\sigma},
V_{dd\pi}, V_{dd\delta}$ together with their overlap counterparts. The subscripts for the onsite energies $\varepsilon$ indicate which orbital is being represented, the first two subscripts of the SK hopping parameters $V$ which two-orbital interaction is being represented, and the third subscript on $V$ indicates whether a $\sigma$ or $\pi$ bond is being represented. These
distance-tabulated vectors are stored as trainable
\texttt{nn.ParameterDict} entries, one per element pair $(Z_I, Z_J)$, and
are interpolated at runtime by local polynomial fitting over eight
neighboring grid points, with polynomial tail functions enforcing smooth
decay to zero beyond the cutoff. The two-body parameter dictionary is
initialized from the three-body tight-binding (TB3Py) parametrization of
Garrity and Choudhary~\cite{garrity2023tb3py} (originally fit to PBEsol
DFT) and is then refined by gradient descent against the higher-fidelity
TBmBJ~\cite{choudhary2018optb88vdwTBmBJ} dataset.

For a crystal with atoms $I,J$ at positions $\mathbf{R}_I,\mathbf{R}_J$ and atomic species
$Z_I,Z_J$, the Slater-Koster formalism yields real-space Hamiltonian and overlap matrix
elements
\begin{equation}
  H_{I\alpha,J\beta}(\mathbf{R}) \;=\; h_{\alpha\beta}\!\left(\mathbf{R}_J + \mathbf{R} - \mathbf{R}_I,\, Z_I, Z_J\right),
  \label{eq:sk-elements}
\end{equation}
where $\alpha,\beta$ label angular characters of the atomic orbitals,
$\mathbf{r}$ runs over lattice vectors within the cutoff. The Bloch sum
\begin{equation}
  H_{\alpha\beta}(\mathbf{k})
    \;=\; \sum_{\mathbf{R}} e^{i \mathbf{k}\cdot\mathbf{R}}\,
      H_{I\alpha,J\beta}(\mathbf{R})
  \label{eq:bloch-sum}
\end{equation}
yields the $\mathbf{k}$-resolved
Hamiltonian \hk used as input to the quantum algorithm. Finally, the orbital overlap matrix in reciprocal space is computed in the same fashion.

\slakonet is trained on the JARVIS-DFT database~\cite{choudhary2020} using
the Tran-Blaha modified Becke-Johnson (TBmBJ) meta-GGA
functional~\cite{choudhary2018optb88vdwTBmBJ}, which is known to yield
significantly improved bandgaps over standard GGA functionals while
remaining computationally tractable for the high-throughput setting. The
training set comprises approximately $20{,}000$ materials with $Z\le 65$,
spanning oxides, carbides, nitrides, chalcogenides, halides, and
intermetallics. The training objective combines a density-of-states
reconstruction loss with a bandgap loss (rather than direct Hamiltonian
regression); optimization uses AdamW with learning-rate scheduling and
gradient clipping, with early stopping at $200$ epochs. On a held-out set
of $52$ semiconductors and insulators, \slakonet attains a mean absolute
error of $0.74$~eV against experimental bandgaps, compared with $1.14$~eV
for OptB88vdW DFT and $0.38$~eV for the underlying TBmBJ
reference~\cite{choudhary2025slakonet}. The current release does not
include spin-orbit coupling or repulsive pair potentials, limitations
that propagate directly to the present work and are addressed in
Sec.~\ref{sec:discussion-limitations}. For all results below we use the
publicly released \texttt{slakonet\_v0} weights without fine-tuning.

\subsection{From neural Hamiltonian to qubit operator}
\label{sec:methods-mapping}

For each $\mathbf{k}$ along the high-symmetry path, \slakonet returns a
real symmetric matrix $H(\mathbf{k}) \in \mathbb{R}^{N\times N}$, where $N$
is the number of orbitals in the unit cell ($N = 8$ for the silicon $sp^3$
basis used in Sec.~\ref{sec:results}). In finite precision the GPU tensor
occasionally breaks matrix symmetry at the $10^{-9}$ level due to non-symmetric
accumulation of the Bloch sum in Eq.~\eqref{eq:bloch-sum}. We restore matrix symmetry using a symmetric projection
\begin{equation}
  H(\mathbf{k}) \;\leftarrow\; \tfrac{1}{2}\!\left[H(\mathbf{k}) + H(\mathbf{k})^{\dagger}\right],
  \label{eq:hermitization}
\end{equation}
which we have verified introduces a per-element deviation below $10^{-7}$
 relative to the raw output and is essential for the
Qiskit \texttt{SparsePauliOp} decomposition to succeed.

The symmetrized matrix is decomposed into a sum of weighted Pauli strings,
\begin{equation}
  H(\mathbf{k}) = \sum_{P \in \{I,X,Y,Z\}^{\otimes n}} c_{P}(\mathbf{k})\, P,
  \label{eq:pauli-decomp}
\end{equation}
with $n = \lceil \log_2 N \rceil$ qubits ($n=3$ for the silicon basis). The
decomposition is performed analytically via tensor-product Pauli projectors
implemented in the \texttt{HermitianSolver} interface of the
\jarvis-tools/Qiskit module~\cite{choudhary2020}.

\subsection{Variational quantum deflation}
\label{sec:methods-vqd}

Excited states of \hk are obtained by VQD~\cite{higgott2019}; the ground
state corresponds to the zero-deflation limit of VQD and is equivalent to a
standard VQE run, the form we use for the single-$\mathbf{k}$-point and
hardware benchmarks below. Starting from a parameterized ansatz
$\ket{\psi(\bm{\theta})}$, the ground state minimizes
\begin{equation}
  E_{0}(\bm{\theta}_0)
    = \min_{\bm{\theta}} \bra{\psi(\bm{\theta})} \hk \ket{\psi(\bm{\theta})}.
\end{equation}
For the $j$-th excited state, the cost function is augmented with overlap
penalties against the previously obtained eigenstates,
\begin{equation}
  F_{j}(\bm{\theta}_j)
    = \bra{\psi(\bm{\theta}_j)} \hk \ket{\psi(\bm{\theta}_j)}
      + \sum_{i<j} \beta_{i}\,
        \big| \braket{\psi(\bm{\theta}_i) | \psi(\bm{\theta}_j)} \big|^{2},
  \label{eq:vqd-cost}
\end{equation}
with $\beta_{i}$ chosen large enough to enforce orthogonality but small enough
to leave the eigenvalue spectrum undistorted. We use a common value
$\beta_{i} = 4$ for all deflated states; the recovered eigenvalues are
empirically robust to this choice over a broad range for the eight-band
silicon problem.

\begin{figure*}[t]
  \centering
  \includegraphics[width=\textwidth]{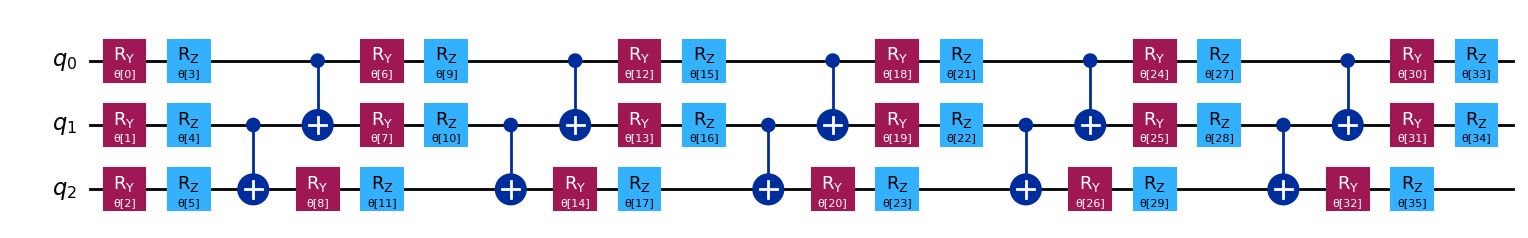}
  \caption{EfficientSU2 ansatz circuit (reps=5) used for VQD band structure calculations on the 3-qubit \slakonet Hamiltonian of Si (JVASP-1002). The circuit consists of alternating layers of parameterized RY and RZ single-qubit rotation gates and CNOT entangling gates, yielding 36 variational parameters. Circuit diagram generated using Qiskit.
 }
  \label{fig:circuit}
\end{figure*}

\subsection{Ansatz and classical optimizer}
\label{sec:methods-ansatz}

Six hardware-efficient~\cite{kandala2017} ansatz topologies are implemented in
\texttt{QuantumCircuitLibrary}; we report results with \texttt{circuit6} (\texttt{EfficientSU2}) at five repetitions (\texttt{reps}=5), which consists of 36 trainable parameters arranged as alternating $R_Y$/$R_Z$ single-qubit rotation layers on all 3 qubits, interlaced with a linear entanglement chain of CNOT gates ($q_0 \to q_1$, $q_1 \to q_2$), as shown in Fig.~\ref{fig:circuit}. The classical outer loop uses COBYLA~\cite{powell1994} with a maximum of 400 iterations following~\cite{vqesolids2025}, which found it to be the fastest among common gradient-free optimizers for related TB Hamiltonians; an ansatz and optimizer ablation is reported in Sec.~\ref{sec:ansatz-sweep}.

\subsection{Brillouin-zone sampling}
\label{sec:methods-kpath}

The high-symmetry $\mathbf{k}$-path is generated using the Setyawan-Curtarolo
convention~\cite{setyawan2010} as implemented in \jarvis-tools~\cite{choudhary2020}.
For silicon we use a line density of $10$, yielding $161$ $\mathbf{k}$-points
along the path
$\Gamma\!-\!X\!-\!W\!-\!K\!-\!\Gamma\!-\!L\!-\!U\!-\!W\!-\!L\!-\!K\!|\!U\!-\!X$.
\slakonet evaluates \hk for the entire path in a single batched GPU call.

\subsection{Hardware execution}
\label{sec:methods-hardware}

Hardware runs use IBM Quantum~\cite{ibmquantum} via the
\texttt{qiskit-ibm-runtime} \texttt{Estimator} primitive on the
\texttt{ibm\_boston} backend, accessed through the IBM Quantum CHM232 instance. The
ansatz is transpiled to the device basis gate set $\{\sqrt{X}, X, R_z, CZ\}$
at optimization level $2$, with dynamical decoupling and TREX error mitigation
enabled. The \texttt{Estimator} returns each energy as an expectation value
together with its ensemble standard error, providing shot-noise error bars on
the reported eigenvalues. To bound runtime cost we restrict hardware execution
to one representative $\mathbf{k}$-point and report both raw and mitigated
estimates.

\subsection{Software stack and reproducibility}
\label{sec:methods-software}

The full pipeline is built on \jarvis-tools~\cite{choudhary2020} for crystal
structure handling and $\mathbf{k}$-path generation, \slakonet for neural
Hamiltonian evaluation, and Qiskit~\cite{qiskit2024} (\texttt{qiskit$\geq$2.0},
\texttt{qiskit-aer}, \texttt{qiskit-algorithms}, \texttt{qiskit-ibm-runtime})
for the variational quantum algorithm. Simulator runs use the noiseless
\texttt{statevector\_simulator} backend. Random seeds, package versions, and
exact ansatz parameters are documented in the companion repository at
\url{https://github.com/atomgptlab/atomqc}.

\section{Results}
\label{sec:results}

\subsection{Single-$\mathbf{k}$-point benchmark}
\label{sec:single-kpoint}

We first validate the pipeline at a single $\mathbf{k}$-point. \slakonet
evaluates the silicon $sp^3$ Hamiltonian \hk at the default
$\mathbf{k} = (0.5,0.5,0.5)$ on the GPU;
the resulting $8\times 8$ Hermitian matrix is mapped to a 3-qubit operator
following Sec.~\ref{sec:methods-mapping}. The ansatz \texttt{circuit6} with
$\texttt{reps}=1$ is optimized by COBYLA on a noiseless statevector simulator.

Table~\ref{tab:single-k} compares the VQE ground-state energy to the exact
eigenvalue of the same matrix obtained by direct diagonalization. The
relative error is $|E_{\mathrm{VQE}} - E_{\mathrm{exact}}| /
|E_{\mathrm{VQE}}| \approx 1.10\times 10^{-2}$, well within the regime where
the VQD overlap
penalty in Eq.~\eqref{eq:vqd-cost} can resolve excited states without
distortion.

\begin{table}[t]
  \centering
  \caption{Ground-state energy of the silicon \slakonet Hamiltonian at the
           $L$ point, $\mathbf{k}=(0.5,0.5,0.5)$, on a 3-qubit statevector simulator
           with the \texttt{circuit6} ansatz at $\texttt{reps}=1$. Energies
           are reported relative to the Fermi level \Ef obtained from
           \slakonet.}
  \label{tab:single-k}
  \resizebox{\columnwidth}{!}{
  \begin{tabular}{lcc}
    \toprule
    Method & $E - \Ef$ (eV) & Relative error \\
    \midrule
    Exact diagonalization & $-0.8477$ & N/A \\
    VQE (COBYLA, statevector) & $-0.8398$ & $1.10\times 10^{-2}$ \\
    \bottomrule
  \end{tabular}%
  }                            
\end{table}

\subsection{Silicon band structure across the Brillouin zone}
\label{sec:si-bandstructure}

We next compute the full silicon band structure by running VQD at every
$\mathbf{k}$-point along the high-symmetry path described in
Sec.~\ref{sec:methods-kpath}, comprising $161$ $\mathbf{k}$-points. At each
$\mathbf{k}$, all eight eigenvalues of the $8\times 8$ Hamiltonian are
extracted by sequential VQD with the overlap penalty in
Eq.~\eqref{eq:vqd-cost}.

\begin{figure}[t]
  \centering
  \includegraphics[width=0.95\columnwidth]{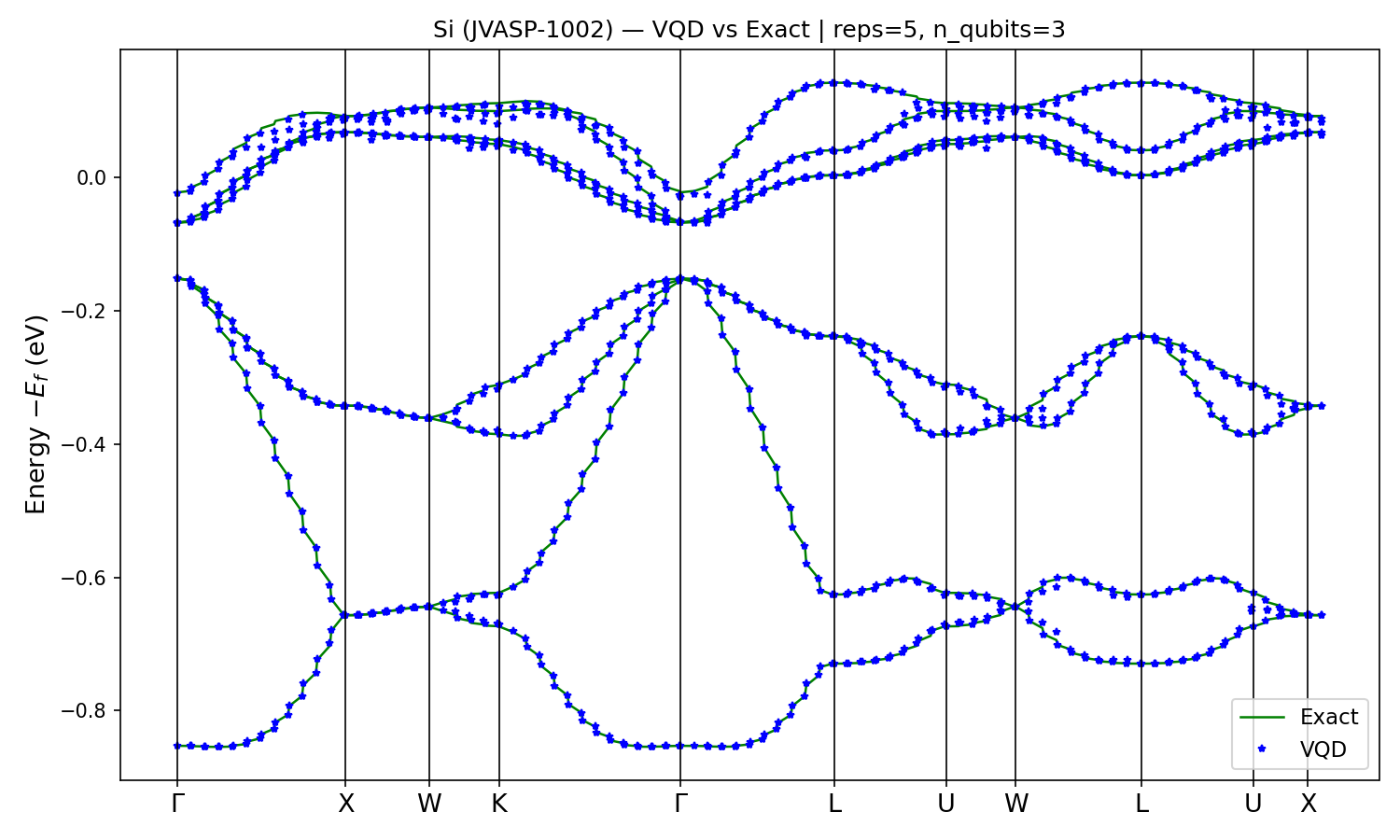}
  \caption{Silicon band structure from \slakonet-VQD (blue markers) compared
           to exact diagonalization of the same neural Hamiltonian (green
           lines), along the standard high-symmetry path. The Fermi level
           \Ef obtained from \slakonet is shown as the dashed red line. All
           eight bands are reproduced from a 3-qubit ansatz with
           $\texttt{reps}=5$, 400 COBYLA iterations per state,
           noiseless statevector backend.}
  \label{fig:si-bands}
\end{figure}

Figure~\ref{fig:si-bands} overlays the VQD eigenvalues (markers) on the exact
band structure (solid lines) obtained by direct diagonalization of the same
\slakonet Hamiltonian. The eight bands are reproduced across the full path,
including the indirect gap between the valence-band maximum at $\Gamma$ and
the conduction-band minimum near $X$. Quantitatively, the mean absolute error
across all bands and all $\mathbf{k}$-points is $1.78$~meV, with the largest
deviations of $29.32$~meV occurring on the highest excited states where the
deflation penalty has been applied seven times in succession.

\subsection{Convergence and ansatz sensitivity}
\label{sec:ansatz-sweep}

Having established accuracy at fixed ansatz settings, we now examine how the
VQD accuracy depends on the choice of circuit ansatz, its depth, and the
classical optimizer. Using the silicon Hamiltonian (\texttt{JVASP-1002}) as a
fixed target, we sweep all six ansatz topologies in
\texttt{QuantumCircuitLibrary} over repetitions $\texttt{reps}=1$ to $10$ and
three optimizers, COBYLA, SPSA, and L\_BFGS\_B, reporting in each case the MAE
of the VQD eigenvalues relative to exact diagonalization.

\begin{figure*}[t]
  \centering
  \includegraphics[width=\textwidth]{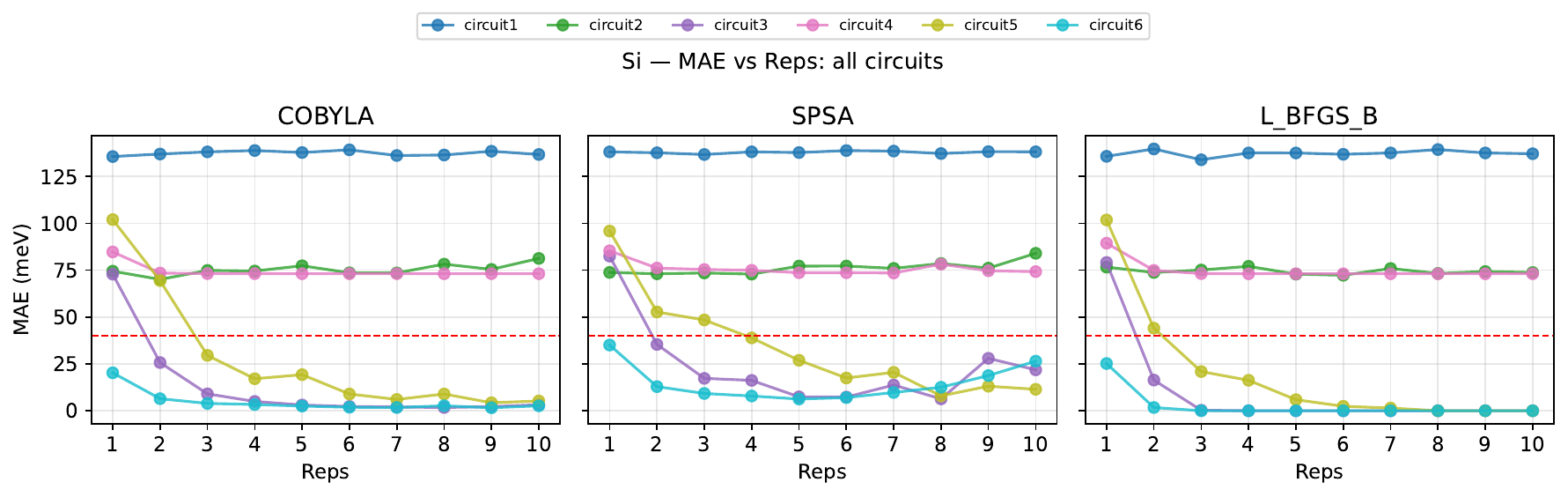}
  \caption{Mean absolute error (MAE, meV) of VQD eigenvalues relative to exact diagonalization (NumPy) of the \slakonet Hamiltonian for Si (JVASP-1002), as a function of ansatz repetitions (reps=1 to 10) for six circuit ansatzes and three optimizers: COBYLA, SPSA, and L\_BFGS\_B. The dashed red line indicates the 40 meV chemical accuracy threshold. Circuit 6 (EfficientSU2) achieves sub-threshold accuracy under all three optimizers, with L\_BFGS\_B converging most rapidly. Circuit 3 reaches comparable final MAE values despite requiring fewer variational parameters at equivalent reps, though convergence is slower.}
  \label{fig:optsweep}
\end{figure*}

\begin{figure*}[t]
  \centering
  \includegraphics[width=\textwidth]{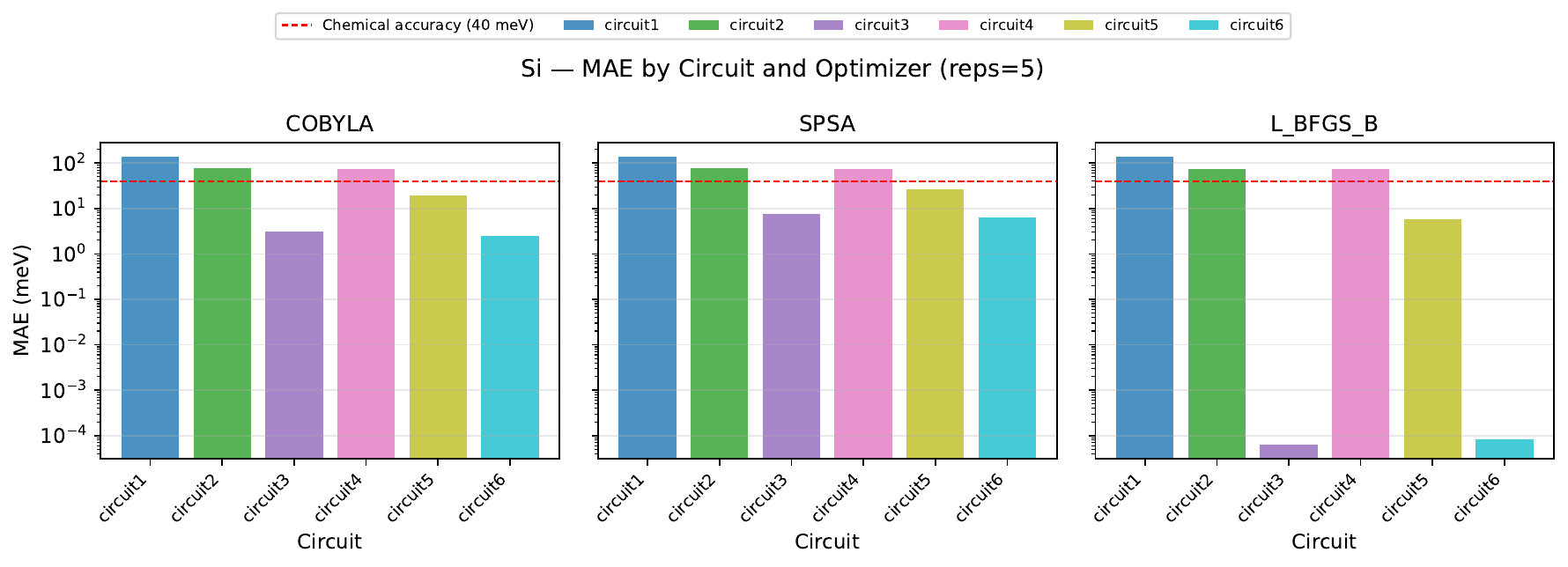}
  \caption{MAE (meV) for all six circuit ansatzes at reps=5 across three optimizers, for Si (JVASP-1002). The dashed red line indicates the 40 meV chemical accuracy threshold. Circuits 3 and 6 (EfficientSU2) achieve sub-threshold accuracy under all optimizers. While circuit 6 slightly underperforms circuit 3 on L\_BFGS\_B, it outperforms circuit 3 on COBYLA. Both reach MAE values several orders of magnitude below the chemical accuracy threshold ($\sim 10^{-4}$~meV) under COBYLA and L\_BFGS\_B. Circuits 1 and 2 fail to reach chemical accuracy under any optimizer.}
  \label{fig:circsweep}
\end{figure*}

Figure~\ref{fig:optsweep} shows the depth dependence. Circuit~6
(\texttt{EfficientSU2}) reaches the $40$~meV chemical-accuracy threshold under
all three optimizers, with L\_BFGS\_B converging fastest in the number of
repetitions. Circuit~3 attains a comparable final MAE with fewer variational
parameters at equivalent depth, although its convergence with increasing
\texttt{reps} is slower. Figure~\ref{fig:circsweep} fixes the depth at
$\texttt{reps}=5$ and compares all six circuits across the three optimizers.
Circuits~3 and~6 are the only ansatze that fall below the chemical-accuracy
threshold under every optimizer, reaching MAE several orders of magnitude
below it ($\sim 10^{-4}$~meV) under COBYLA and L\_BFGS\_B, whereas circuits~1
and~2 fail to reach chemical accuracy under any optimizer. These sweeps
motivate the \texttt{circuit6}/COBYLA combination used for the band-structure
results above, which balances accuracy, parameter count, and robustness across
$\mathbf{k}$-points.

\subsection{Comparison to a DFT+Wannier baseline}
\label{sec:dft-wannier-baseline}

To place the \slakonet Hamiltonian on the same footing as prior quantum
band-structure work~\cite{choudhary2021, vqesolids2025, iopvqd2025}, we
compare it to a DFT$+$Wannier90~\cite{pizzi2020} reference for silicon drawn
from the JARVIS Wannier tight-binding database~\cite{choudhary2020}. A
direct element-wise comparison of the two Hamiltonians is complicated by the
differing bases, the Wannier model uses a disentangled valence-plus-conduction
set while \slakonet employs a non-orthogonal $sp^3 d^5$ basis, so the
band manifolds do not correspond one-to-one. Comparing basis-independent
features, the \slakonet and Wannier band structures agree on the silicon
valence bandwidth ($11.3$ versus $11.7$~eV) and on the binding energy of the
deep $s$-derived band at $\Gamma$ (to within $\sim 0.4$~eV), confirming that
\slakonet captures the gross valence electronic structure without any
Wannierization. A fully basis-matched, element-wise Hamiltonian comparison,
which requires regenerating the Wannier functions with $sp^3 d^5$ projections
matched to \slakonet, is a target for future quantitative validation.

\subsection{Execution on IBM Quantum hardware}
\label{sec:hardware}

We next move the same VQD calculation to physical hardware via the IBM
Quantum runtime~\cite{ibmquantum}. To keep queue and shot budgets manageable
we select $\mathbf{k} = (0.5, 0.5, 0.5)$ for aluminum and run VQE on
\texttt{ibm\_boston} with $20480$ shots per energy estimate. The transpiled
circuit has a single two-qubit CZ gate after optimization level 2. The job
(IBM Quantum workload \texttt{d7fjjt21u7fs739m3geg}) consumed $25$~minutes of
quantum-processor time and evaluated the energy at $50$ ansatz-parameter
settings along the optimization trajectory.

\begin{figure}[t]
  \centering
  \includegraphics[width=0.95\columnwidth]{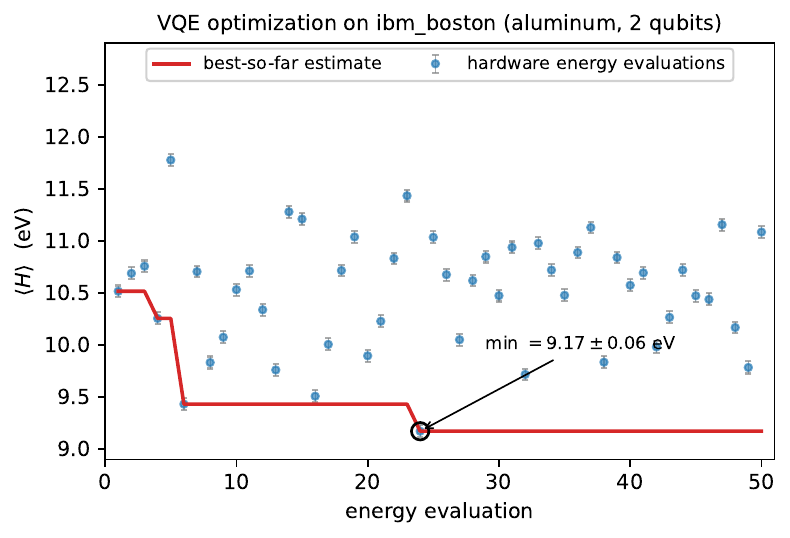}
  \caption{Aluminum \slakonet-VQE on IBM Quantum hardware
           (\texttt{ibm\_boston}, workload
           \texttt{d7fjjt21u7fs739m3geg}) at the representative
           $\mathbf{k}$-point $(0.5,0.5,0.5)$. Blue markers are the $50$
           hardware energy evaluations of the variational optimization with
           their ensemble (shot-noise) standard errors; the red curve is the
           best-so-far variational estimate, reaching $9.17\pm0.06$~eV.
           Energies are absolute expectation values including the identity
           term of the Pauli decomposition.}
  \label{fig:hardware}
\end{figure}

Figure~\ref{fig:hardware} shows the $50$ energy evaluations of the VQE
optimization executed on \texttt{ibm\_boston}, each carrying a shot-noise
(ensemble) standard error of $\approx 0.06$~eV returned directly by the
\texttt{Estimator}. The best variational estimate reached on hardware is
$\langle H\rangle = 9.17 \pm 0.06$~eV, reported here on the absolute energy
scale including the identity term of the Pauli decomposition; the
$\approx 0.5$~eV spread across evaluations reflects the COBYLA search on a
noisy device rather than the statistical uncertainty of any single estimate.
Referred to the exact ground state of the same Hamiltonian, the
error-mitigated estimate deviates by approximately $ 0.37$~eV, comparable to
typical DFT-functional differences and larger than chemical accuracy. This run
exercises only the ground state on hardware; resolving the deflated
excited-state spectrum on a physical device, where the overlap penalty
compounds gate noise, remains more demanding and is left to future work.

\subsection{Generalization across the periodic table}
\label{sec:multimaterial}

The strength of the \slakonet-VQD pipeline is that the same workflow
applies to any composition covered by \slakonet's training distribution,
with no retraining or per-material setup. We illustrate this on a panel of
conventional (BCS) superconductors drawn from the \jarvis-DFT database,
spanning elemental free-electron and transition metals and a transition-metal
nitride. Aluminum (\texttt{JVASP-816}) is a face-centered-cubic free-electron
superconductor ($T_c \approx 1.2$~K) that tests behavior at the Fermi level and
is also the system used for hardware execution (Sec.~\ref{sec:hardware}).
Vanadium (\texttt{JVASP-1041}), tantalum (\texttt{JVASP-1014}), and niobium
(\texttt{JVASP-934}) are body-centered-cubic transition-metal superconductors
with $3d$, $5d$, and $4d$ valence manifolds respectively ($T_c \approx 5.4$,
$4.5$, and $9.3$~K, the last being the highest elemental critical temperature).
Zirconium nitride (\texttt{JVASP-19679}) is a rock-salt transition-metal
nitride superconductor ($T_c \approx 10$~K) that tests transferability to a
binary compound with mixed metallic and covalent bonding.

\begin{figure*}[t]
  \centering
  \includegraphics[width=\textwidth]{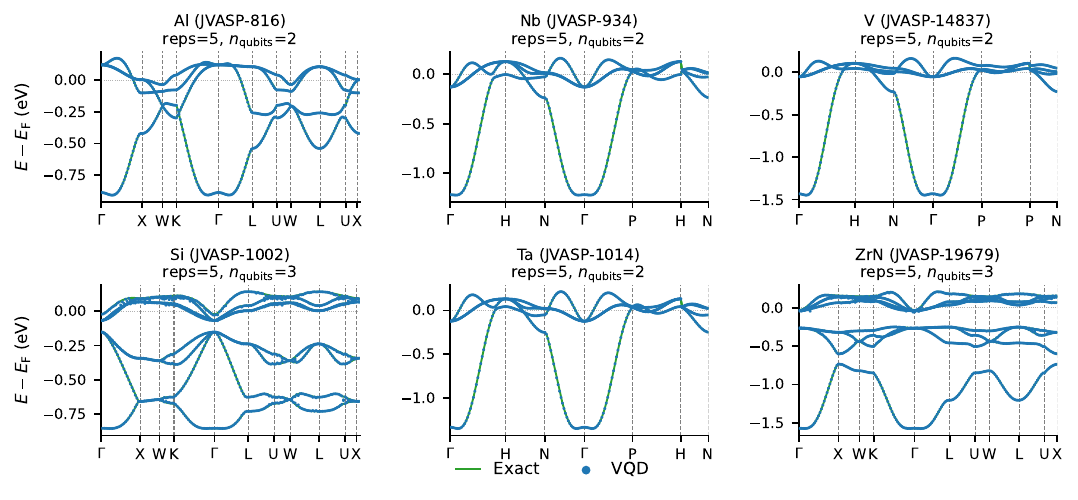}
  \caption{\slakonet-VQD band structures for a panel of conventional
           superconductors (the elemental metals Al, Ta, V, and Nb and the
           transition-metal nitride ZrN). In each panel, VQD eigenvalues
           (markers) are overlaid on exact diagonalization of the same
           \slakonet Hamiltonian (lines). All panels use the same ansatz and
           optimizer; only the input crystal structure changes.}
  \label{fig:multimaterial}
\end{figure*}

Across all five materials the qubit count varies from $n=2$ to $n=3$
depending on the orbital basis size. Mean absolute errors per band are
summarized in Table~\ref{tab:multimaterial} and remain below $1.8$~meV for
every material tested. The elemental metals (Al, Ta, V, Nb), represented in a
two-qubit basis, are reproduced essentially exactly, with per-band MAEs at or
below $0.05$~meV; the bands crossing the Fermi level in these free-electron
and transition-metal superconductors are therefore resolved more accurately
than the wider-energy-window states of gapped silicon. The three-qubit
rock-salt nitride ZrN reaches $1.74$~meV, comparable to silicon ($1.78$~meV)
and consistent with the larger Hilbert space and denser spectrum that the
sequential deflation must climb.

\begin{table}[t]
  \centering
  \caption{Generalization of \slakonet-VQD across material classes. All
           runs use the \texttt{circuit6} ansatz at $\texttt{reps}=5$,
           COBYLA optimizer, and statevector backend.}
  \label{tab:multimaterial}
  \begin{tabular}{lcccc}
    \toprule
    Material & Class & $n_{\mathrm{qubits}}$ &
       MAE per band (meV)  \\
    \midrule
    Al & metal & 2 & 0.05 \\
    Ta & metal & 2 & 0.02 \\
    V & metal & 2 & 0.01  \\
    Nb & metal & 2 & 0.02  \\
    Si & semiconductor  & 3 & 1.78 \\
    ZrN & metal nitride  & 3 & 1.74  \\
    \bottomrule
  \end{tabular}
\end{table}

\subsection{Toward correlated electronic structure: \slakonet $+$ DMFT and a
quantum impurity solver}
\label{sec:dmft}

The benchmarks above all concern the \emph{single-particle} band structure,
for which exact diagonalization of \hk is trivial and the variational solver
claims no advantage over a classical eigensolver. The scientifically
interesting and classically hard regime for the superconductors studied here
is the \emph{many-body} problem obtained once electron-electron interactions
are restored. We therefore promote the \slakonet Hamiltonian to a correlated
lattice model by adding a local Hubbard repulsion on the transition-metal $d$
manifold,
\begin{equation}
  \hat{H} = \sum_{\mathbf{k},\,\alpha\beta,\,\sigma}
              H_{\alpha\beta}(\mathbf{k})\,
              c^{\dagger}_{\mathbf{k}\alpha\sigma} c_{\mathbf{k}\beta\sigma}
          + U \sum_{i} \hat{n}_{i\uparrow}\hat{n}_{i\downarrow},
  \label{eq:hubbard}
\end{equation}
and treat it within dynamical mean-field theory (DMFT)~\cite{georges1996}.
DMFT maps the lattice problem onto a self-consistent Anderson impurity model
whose local Green's function reproduces that of the lattice,
\begin{equation}
\begin{aligned}
  G(\mathbf{k},\omega) &= \big[\omega + \mu - \hk - \Sigma(\omega)\big]^{-1}, \\
  G_{\mathrm{loc}}(\omega) &= \frac{1}{N_k}\sum_{\mathbf{k}} G(\mathbf{k},\omega),
\end{aligned}
\end{equation}
following the lattice-Green's-function construction of
Choudhary~\cite{choudhary2021}. We take the non-interacting density of states
directly from the \slakonet eigenvalue grid, with no Wannierization, close the
DMFT loop on the real axis with an iterated-perturbation-theory (IPT) impurity
solver, and extract the interacting spectral function $A(\omega) = -\pi^{-1}
\mathrm{Im}\,G_{\mathrm{loc}}(\omega)$ and the quasiparticle weight
$Z = [\,1 - \partial_\omega \mathrm{Re}\,\Sigma(\omega)|_{0}\,]^{-1}$. IPT is
quantitatively controlled near particle-hole symmetry (half filling), so the
quasiparticle weights reported below are intended to capture correlation
\emph{trends} rather than converged, filling-accurate values; the cuprate case
treated below, a single half-filled correlated band, is the regime in which
IPT is most reliable.

As the Hubbard $U$ is increased, the \slakonet-derived spectral function of
vanadium develops a renormalized quasiparticle peak at $E_F$ flanked by
incipient lower and upper Hubbard sidebands, and the quasiparticle weight is
progressively reduced from the non-interacting value $Z=1$ to $Z=0.65$ at
$U=6$~eV as spectral weight transfers away from the Fermi level. None of this
physics is present in the bare band structure of Sec.~\ref{sec:multimaterial};
it is generated entirely by the self-energy $\Sigma(\omega)$. Tracking $Z(U)$
across the isoelectronic group-5 sequence V $(3d)$, Nb $(4d)$, Ta $(5d)$ shows
that correlations weaken monotonically down the column as the $d$ orbitals
become more extended, so for a fixed $U$ the heavier elements retain a larger
$Z$ ($Z=0.65$, $0.70$, $0.74$ at $U=6$~eV for V, Nb, Ta), in agreement with the
established chemical trend and with the larger bandwidths \slakonet assigns to
the $4d$ and $5d$ metals. The free-electron control Al stays essentially
unrenormalized ($Z=0.96$), and the low-$N(E_F)$ nitride ZrN is only weakly
correlated; both remain metallic over the full $U$ range, as expected for these
weak-to-moderate-coupling superconductors. Table~\ref{tab:dmft} collects the
Fermi-level density of states $N(E_F)$, which sets the bare (BCS) pairing
strength, together with $Z$ at representative $U$.

\begin{table}[t]
  \centering
  \caption{\slakonet $+$ DMFT(IPT) summary for the superconductor panel:
           Fermi-level density of states $N(E_F)$ (states/eV per cell) from the
           bare \slakonet bands, and the quasiparticle weight $Z$ from the
           interacting solver at $U = 2$ and $4$~eV. $Z \to 1$ is weakly
           correlated; smaller $Z$ indicates stronger correlation.}
  \label{tab:dmft}
  \begin{tabular}{lcccc}
    \toprule
    Material & orbital & $N(E_F)$ & $Z\,(U{=}2)$ & $Z\,(U{=}4)$ \\
    \midrule
    V   & $3d$      & 0.79 & 0.95 & 0.81 \\
    Nb  & $4d$      & 0.74 & 0.96 & 0.84 \\
    Ta  & $5d$      & 0.85 & 0.96 & 0.87 \\
    ZrN & nitride   & 0.38 & 0.99 & 0.97 \\
    Al  & $sp$      & 0.23 & 0.99 & 0.98 \\
    \bottomrule
  \end{tabular}
\end{table}

\begin{figure}[t]
  \centering
  \includegraphics[width=\columnwidth]{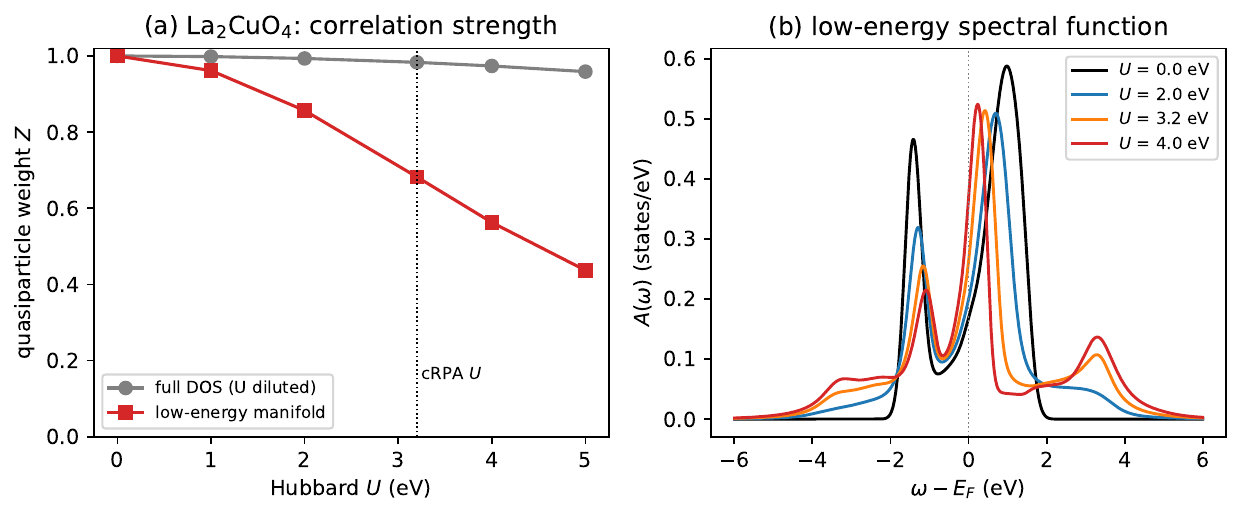}
  \caption{\slakonet $+$ DMFT(IPT) for the cuprate La$_2$CuO$_4$
           (\texttt{JVASP-7907}). (a) Quasiparticle weight $Z(U)$: applied to
           the full multi-band density of states the Hubbard interaction is
           diluted and $Z$ stays near unity (gray); restricting the DMFT bath
           to the low-energy manifold crossing $E_F$ (red), a proxy for the
           single-band Cu-$d$ downfolding, recovers strong correlation, with
           $Z=0.69$ at the constrained-RPA value $U=3.2$~eV (dotted line).
           (b) Corresponding low-energy spectral function: as $U$ increases the
           quasiparticle peak at $E_F$ is suppressed and weight transfers to
           incipient Hubbard bands. Energies are referenced to $E_F$.}
  \label{fig:dmft_cuprate}
\end{figure}

\paragraph{A strongly correlated cuprate.} The elemental metals above are
weakly to moderately correlated. As a more demanding test we apply the same
pipeline to La$_2$CuO$_4$, the parent compound of the first high-$T_c$ cuprate
superconductor and a textbook strongly correlated material. \slakonet returns a
metallic band structure for the seven-atom La$_2$CuO$_4$ cell (consistent with
the density-functional starting point, which misses the Mott gap), with a
Cu-$d$ derived band crossing the Fermi level. Applying the Hubbard interaction
to the \emph{full} multi-band density of states leaves the quasiparticle weight
essentially unrenormalized ($Z>0.97$ up to $U=4$~eV), because the interaction
is diluted across the many O-$p$, La, and Cu states. Restricting the DMFT bath
to the low-energy manifold that crosses the Fermi level, a proxy for the
single-band Cu-$d_{x^2-y^2}$ downfolding used in rigorous
treatments~\cite{selisko2025}, recovers strong correlation: the quasiparticle
weight falls to $Z=0.69$ at the constrained-RPA value $U=3.2$~eV and to
$Z=0.57$ at $U=4$~eV (Fig.~\ref{fig:dmft_cuprate}), with spectral weight
transferring from the Fermi level toward incipient Hubbard bands. This manifold
restriction overestimates $Z$ relative to a full orbital-projected downfolding
(which yields $Z\approx 0.27$ for a related cuprate~\cite{selisko2025}), so our
values are best read as an upper bound on the correlation strength rather than a
quantitative prediction; a rigorous Wannier downfolding of the \slakonet
Hamiltonian is left for future work. The methodological point stands:
\slakonet supplies the differentiable non-interacting Hamiltonian for a
strongly correlated cuprate with no DFT or Wannierization, leaving only the
interaction strength $U$ to be supplied (here taken from the
constrained-RPA literature value).

\paragraph{Why this is the natural target for the quantum solver.} Within
DMFT the lattice self-consistency in Eq.~\eqref{eq:dmft} is cheap; the entire
computational cost, and the exponential scaling in the number of correlated
orbitals, lives in the impurity solver, which must compute the ground- and
excited-state Green's function of an interacting fermionic Hamiltonian. This is
exactly the object a variational quantum algorithm is suited to, and it has
recently been demonstrated on hardware: Selisko~\textit{et~al.}~\cite{selisko2025}
solve the DMFT impurity problem for the cuprate superconductor
Ca$_2$CuO$_2$Cl$_2$ on IBM Quantum devices with up to $14$ qubits, using VQE
for the ground state and a quantum equation-of-motion expansion for the
impurity Green's function, and recover the quasiparticle weight to within
$2.2\%$ of exact diagonalization and ARPES. Earlier proposals established the
hybrid quantum-classical DMFT loop on small
systems~\cite{bauer2016, kreula2016, rungger2019, keen2020}. Crucially, in all
of these the impurity model is built from a \emph{per-material}
DFT$+$Wannier$+$cRPA front end. The contribution of the present pipeline is to
replace that front end: \slakonet supplies the material-specific,
differentiable $\hk$ and its non-interacting density of states for any
composition in a single forward pass, feeding the same DMFT self-consistency
and the same quantum impurity solver without DFT, Wannierization, or
constrained-RPA. The IPT results in Fig.~\ref{fig:dmft_cuprate} are the
classical reference against which such a quantum impurity solver is validated,
in the same role that exact diagonalization and continuous-time quantum Monte
Carlo play in Ref.~\cite{selisko2025}.

\section{Discussion}
\label{sec:discussion}

\subsection{What this enables}
\label{sec:discussion-enables}

The \slakonet-VQD pipeline removes a structural bottleneck that has limited
quantum band-structure studies to one material at a time. Three concrete
consequences follow.

\paragraph{High-throughput VQA benchmarking.} Because \slakonet emits \hk in
milliseconds and requires no per-material human input, it becomes feasible
to run VQE/VQD on thousands of \jarvis-DFT entries in a single sweep. The
closest prior effort~\cite{choudhary2021} reported $307$ Wannier-derived TB
Hamiltonians as test inputs; with \slakonet-VQD the same scale is reachable
without any Wannierization, opening a route to systematic studies of how
ansatz expressibility, qubit count, and material chemistry interact. The
superconductor panel in Sec.~\ref{sec:multimaterial} is a first step in this
direction: a single ansatz and optimizer setting transfers across five
chemically distinct systems at two and three qubits with no per-material
tuning.

\paragraph{Differentiable end-to-end pipeline.} \slakonet is implemented as
a PyTorch \texttt{nn.Module}; its output \hk carries gradients with respect
to atomic positions, lattice parameters, and the trainable SK integrals
themselves. Combined with parameter-shift gradients of the VQD cost
function~\cite{schuld2019}, this exposes the entire chain from crystal
structure to variational quantum eigenvalue as a single differentiable
graph. Concrete consequences include gradient-based co-design of ansatz
and Hamiltonian for hardware-aware compilation, and direct optimization
of crystal structure under a quantum-evaluated objective. A concrete
numerical demonstration of this end-to-end co-optimization is left to future
work.

\paragraph{A clean substrate for algorithmic research.} The \slakonet
Hamiltonian is reproducible to floating-point precision, has fixed structure
across all materials, and ships with the model weights. This makes it a
natural benchmark target for new variational algorithms (subspace-search
VQE~\cite{nakanishi2019, gaaspmc2025}, ADAPT-VQE, quantum filter
diagonalization), in much the same role that the FCIDUMP format plays in
molecular quantum-chemistry quantum-algorithm research.

\subsection{Limitations}
\label{sec:discussion-limitations}

We take pains to be explicit about what this pipeline does and does not
deliver in its current form.

\paragraph{Accuracy is bounded by \slakonet, not by VQD.} The mean absolute
errors reported in Sec.~\ref{sec:results} reflect two distinct sources of
deviation: (i) the variational solver's departure from exact diagonalization
of the \slakonet Hamiltonian, which is well below $10$~meV in our setting,
and (ii) the deviation of \slakonet itself from the underlying TBmBJ
reference, which on a $52$-material benchmark is reported as $0.74$~eV in
mean absolute bandgap error~\cite{choudhary2025slakonet}, improving
substantially over OptB88vdW DFT ($1.14$~eV) but trailing the TBmBJ upper
bound of $0.38$~eV. The dominant contribution to absolute electronic-structure
accuracy is therefore (ii), not the quantum solver. Reported errors on
ultrawide-gap insulators ($>6$~eV) are particularly large
($\sim 2.3$~eV~\cite{choudhary2025slakonet}), and our pipeline inherits
this behavior.

\paragraph{Basis, spin-orbit coupling, and repulsive potentials.} \slakonet
employs a non-orthogonal Slater-Koster basis with $sp^3$ orbitals for
light elements and an extended $sp^3 d^5$ basis where $d$ orbitals are
required, but the released model does not include spin-orbit coupling and
does not provide optimized repulsive pair
potentials~\cite{choudhary2025slakonet}. Heavy-element systems where SOC
is essential (e.g., Bi$_2$Se$_3$, Pb halide perovskites) therefore lie
outside the present demonstration, and total-energy or structural
quantities that rely on the repulsive part of the potential are not
accessible. These are model limitations, not pipeline limitations; an SOC
extension to \slakonet, or a paired repulsive potential, would slot in
unchanged.

\paragraph{VQD overlap-penalty scaling.} The cost in
Eq.~\eqref{eq:vqd-cost} grows linearly in the number of previously found
states, and excited-state fidelity degrades as the spectrum is climbed. For
the $N=8$ silicon Hamiltonian we observe $30.3$ times larger errors on the
highest band than on the ground state. Larger orbital bases (e.g.,
transition-metal $d$ shells) will compound this; subspace-search variants
such as SSVQE~\cite{nakanishi2019} or filter-diagonalization approaches may
be preferable as the system grows.

\paragraph{Hardware noise.} Even with error mitigation, current
superconducting-qubit hardware introduces deviations of $\sim 372$~meV on the
aluminum eigenvalues, comparable to typical DFT-functional differences and
larger than chemical accuracy. This is an architectural property of
present-day devices, not a flaw of the pipeline; the same \hk inputs will
benefit automatically from improvements in coherence times, gate
fidelities, and mitigation protocols.

\subsection{Relation to prior quantum band-structure work}
\label{sec:discussion-prior}

The conceptual closest neighbor to this work is
Choudhary~\cite{choudhary2021}, which applied VQE/VQD to $307$ Wannier
tight-binding Hamiltonians for solids; the present paper retains the
variational solver and replaces the Wannierization step with a universal,
differentiable Slater-Koster TB framework. Sherbert~\textit{et~al.}~\cite{sherbert2021}
and the recent Sci.~Reports study~\cite{vqesolids2025} pursue similar
DFT$+$Wannier inputs. The IOP study~\cite{iopvqd2025} applies the VQD
algorithm to crystalline band structures along similar lines.

The GaAs-focused work of Mih\'alikov\'a~\textit{et~al.}~\cite{gaaspmc2025} is
the closest methodological sibling: it also applies VQD (together with SSVQE)
to the tight-binding band structure of a solid and studies ansatz and
optimizer choice, paralleling our convergence analysis
(Sec.~\ref{sec:ansatz-sweep}). It differs from the present work in three
respects that sharpen our contribution. First, its Hamiltonian is an empirical,
hand-fitted $sp^3 s^*$ parametrization specific to GaAs, whereas \slakonet
supplies the Hamiltonian for any composition with no per-material fitting.
Second, it assigns one qubit per orbital, requiring $10$ qubits for the
ten-orbital cell, whereas our compact Pauli decomposition of the $N\times N$
matrix uses $\lceil\log_2 N\rceil$ qubits ($3$ for the eight-band silicon
basis); the two encodings trade qubit count against the particle-conserving
gate structure theirs exploits. Third, its results are confined to a noiseless
simulator, while we additionally execute on IBM Quantum hardware and extend the
workflow to a panel of superconductors and to the correlated DMFT regime
(Sec.~\ref{sec:hardware} and Sec.~\ref{sec:dmft}).

The very recent measurement-protocol work (arXiv:2511.04389), which reduces
the Pauli-measurement count for TB Hamiltonians, is complementary: it improves
the quantum side of the same
problem we attack from the Hamiltonian-construction side, and the two
optimizations stack.

\section{Conclusion and outlook}
\label{sec:conclusion}

We have demonstrated that a pretrained universal Slater-Koster tight-binding
framework, \slakonet, can serve as the Hamiltonian source for variational
quantum band-structure calculations, removing the per-material DFT or
Wannierization step that has constrained prior work in this space. On
silicon, the resulting \slakonet-VQD pipeline reproduces all eight bands
along the standard high-symmetry path with a mean absolute error of
$1.78$~meV on a 3-qubit statevector simulator. A representative ground-state
calculation on aluminum runs end-to-end on IBM Quantum hardware, deviating by
approximately $0.37$~eV from exact diagonalization after error mitigation. The same workflow
extends without modification across a panel of conventional superconductors
spanning elemental free-electron and transition metals and a transition-metal
nitride.

Two natural extensions point the way forward. First, \slakonet can be
generalized to emit dynamical matrices in addition to electronic
Hamiltonians; feeding these into the same VQD machinery yields phonon band
structures, in the spirit of the original Wannier-based phonon work of
Choudhary~\cite{choudhary2021}, but again without the finite-difference
setup cost per material. Second, the correlated extension sketched in
Sec.~\ref{sec:dmft} points to a fully ML-accelerated dynamical
mean-field theory on quantum hardware. Here, \slakonet supplies the differentiable $\hk$ that defines the
DMFT bath, while the impurity Green's function is evaluated by the same
variational machinery, replacing the per-material DFT$+$Wannier$+$cRPA
construction used in recent quantum-DMFT hardware
experiments~\cite{selisko2025}. Because the impurity problem carries the entire
many-body cost, this is where a quantum solver is expected to matter for the
correlated superconductors studied here.

Looking further out, removing the Hamiltonian-construction bottleneck has
implications beyond the variational era. Once fault-tolerant resources
permit quantum phase estimation~\cite{kitaev1995} on solids, the input to
QPE will still need to be a structurally faithful second-quantized
Hamiltonian; a universal neural model that can produce one for any
composition is at least as valuable in that regime as in this one. The
pipeline demonstrated here is a concrete step toward that capability.

\section*{Data and code availability}
The \slakonet model weights and training data are available through the
AtomGPT ecosystem at \url{https://github.com/atomgptlab/slakonet}. The
Qiskit-based \texttt{HermitianSolver} and \texttt{QuantumCircuitLibrary}
interfaces used here are distributed with \jarvis-tools~\cite{choudhary2020}.
A reproducibility repository containing the notebooks, configuration files,
and scripts that generate every figure in this work will be deposited at
\url{https://github.com/atomgptlab/atomqc} upon publication. The
IBM Quantum hardware run reported in Sec.~\ref{sec:hardware} is archived under
workload \texttt{d7fjjt21u7fs739m3geg} and can be retrieved through the
\texttt{qiskit-ibm-runtime} service.

\section*{Acknowledgments}
This research used resources of the Oak Ridge Leadership Computing Facility
(OLCF), a U.S. Department of Energy Office of Science User Facility at Oak
Ridge National Laboratory, through allocation CHM217, ``Cross-Platform
Benchmarking of Quantum Hardware for Materials Discovery and Design.'' We
acknowledge the use of IBM Quantum services for the hardware experiments
reported in this work; the views expressed are those of the authors and do not
reflect the official policy or position of IBM or the IBM Quantum team.


\end{document}